\documentclass[final,3p,times]{elsarticle}
\usepackage{axodraw}
\usepackage{cancel}
\bibstyle{elsarticle-num}

\def\SOFTSUSY{{\tt SOFTSUSY}}
\def\HSSOFTSUSY{{\tt HidSecSOFTSUSY}}
\def\SUSYB{{$\cancel{\rm SUSY}$}}
\newcommand{\newc}{\newcommand}
\newc{\code}[1]{{\tt #1}}
\journal{Computer Physics Communications}

\begin{document}

\begin{frontmatter}

\title{\HSSOFTSUSY: Incorporating effects from hidden sectors \\in the numerical computation of the
  MSSM spectrum}

\author{Boaz Keren-Zur}
\address{}

\begin{abstract}
\SOFTSUSY~is a software designed to solve the RG equations of the MSSM and compute its low energy spectrum. \HSSOFTSUSY~is an extension of the \SOFTSUSY~package which modifies the beta functions to include contributions from light dynamic fields in the hidden sector.
\end{abstract}

\end{frontmatter}

\section{Program Summary}
\noindent{\em Program title:} \HSSOFTSUSY{}\\
{\em Program obtainable
  from:} {\tt http://www.tau.ac.il/$\sim$kerenzu}\\
{\em Distribution format:}\/ tar.gz\\
{\em Programming language:} {\tt C++}, {\tt fortran}\\
{\em Computer:}\/ Personal computer\\
{\em Operating system:}\/ Tested on GNU/Linux\\
{\em Word size:}\/ 32 bits\\
{\em External routines:}\/ requires an installed version of \SOFTSUSY.\\
{\em Typical running time:}\/ a few seconds per parameter point.\\
{\em Nature of problem:}\/ Calculating supersymmetric particle spectrum and
mixing parameters while incorporating dynamic modes from the hidden sector
into the renormalization group equations. The solution to the equations must be consistent
with a high-scale boundary condition on supersymmetry breaking parameters,
as well as a weak-scale boundary condition on gauge couplings, Yukawa couplings and the Higgs potential parameters.\\
{\em Solution method:}\/ Nested iterative algorithm. \\

\newpage

\section{Introduction}

Supersymmetry is an appealing framework for physics beyond the standard model as it solves the hierarchy problem, leads to gauge coupling unification and provides a basis for a UV completion of the standard model. Supersymmetry is not observed in nature (at least not below the 100 GeV scale) and is assumed to be broken by the dynamics of an additional hidden sector.
In the minimal supersymmetric standard model (MSSM) the hidden sector is integrated out at a scale $M$ and soft SUSY breaking (\SUSYB) terms are introduced into the effective Lagrangian. The value of these effective couplings in the IR can be obtained by evolving the renormalization group equations down from the scale $M$. The package \SOFTSUSY\cite{Allanach:2001kg}~solves this set of coupled differential equations by an iterative process for which the boundaries are the $Z$ boson mass, where the SM parameters are known, and the scale $M$, where the soft \SUSYB~ terms are provided by the user of the software.

This, however, is not necessarily the entire picture. The hidden sector might contain degrees of freedom which are lighter than the scale $M$ and have non-trivial contributions to the beta functions of the MSSM parameters. For example, in models of direct gauge mediation the hidden sector contains pseudomoduli which have masses of the order of the soft \SUSYB~terms, and might be charged under the standard model (e.g. \cite{Zur:2008zg}). The contributions of these fields to the gauge coupling beta function is a leading order effect. Another example is referred to as extra-ordinary gauge mediation (EOGM)\cite{Cheung:2007es}, where the theory has multiple messenger scales. In such theories the various messengers are integrated out at different scales and at the intermediate scales the effective theory contains both soft \SUSYB~terms (generated by the integration out of heavy messengers) and dynamical messengers which contribute at leading order to the RG flow of the gauge couplings and MSSM scalar masses.

In order to incorporate the contributions from the hidden sector into \SOFTSUSY~an 
extension of the package named \HSSOFTSUSY~was written. The input is a list of the relevant superfields in the hidden sector including their masses and representations under the MSSM gauge group. The running of the RG flow is split into several stages defined by the masses of the dynamical hidden sector particles. The beta functions are modified according to the interactions of the dynamical particles at each stage, and between the stages the software takes into account threshold effects.
The extended package includes the following features:

\begin{itemize}
\item
The user provides information regarding the hidden sector fields. Their contribution to the beta function of the MSSM gauge coupling is computed automatically.

\item
EOGM models can be implemented by marking hidden sector fields as messengers. Such fields have additional contributions to the beta functions and important threshold effects which are also computed automatically. The magnitude of these effects depends on the pattern of SUSY breaking and there are several parameters that should be provided by the user.

Since doublet and triplet messengers are introduced separately, models of doublet-triplet splitting are naturally accommodated in the software.

\item
Advanced users may define parameters in the hidden sector and use the $\SOFTSUSY$~engine to compute
their RG flow. This requires some basic knowledge of C++.

\item
More advanced threshold effect can also be introduced using user defined functions.

\end{itemize}

\section{The MSSM RG flow and contributions from hidden fields}
\label{sec_HidSeccontribution}
The contribution of a (non-messenger) charged chiral multiplet $\Psi$ to the beta function of the gauge coupling $g_a$ is given by 
\begin{equation}
\label{eq_beta1}
\Delta\beta_a=N_f\frac{g_a^3}{16\pi^2}I_a(\Psi)~,
\end{equation}
where $N_f$ is the number of flavors (or the degeneracy of the field $\Psi$), and $I_a(\Psi)$ is the Dynkin index of the chiral multiplet $\Psi$.
The beta function for the scalar mass $m_{\phi_i}^2$ is modified by
\begin{equation}
\label{eq_beta2}
\Delta\beta_{m_{\phi_i}^2}=\frac{6}{5}Y_ig_1^2 \rm {Tr}[Y m^2]~,
\end{equation}
where $Y_i$ is the hypercharge, and the trace is over all the charged scalars in the hidden sector.
There are no leading order threshold effects which are relevant to the MSSM parameters as the $\Psi$ multiplet is integrated out. 

If the charged chiral multiplet can be considered as a messenger, 
namely the splitting of the mass squared between the fermions and bosons
is of the order of the SUSY breaking scale, $\sqrt{F}$, then additional contributions become important.
First, the following 2-loop contribution to the beta function of the scalar masses becomes relevant 
\begin{equation}
\label{eq_beta3}
\Delta\beta_{m_{i}^2}=\sum_a\frac{g_a^4}{(16\pi^2)^2}C_a(i)\rm{Str}[I_a(\Psi)\mathcal {M}^2]~,
\end{equation}
where $C_a(i)$ is the casimir of the MSSM scalar field $i$, $\mathcal {M}^2$ is the messenger mass matrix, and $\rm{Str}[\ldots]$ is a supertrace. In the case of a messenger, the supertrace of the mass matrix is large enough to make this 2-loop effect important\cite{Cheung:2007es}.

When a messenger $\Psi$ is integrated out, there are important threshold effects.
The MSSM gauginos and scalar masses receive the following contributions
\begin{eqnarray}
\label{eq_threshold}
\Delta M_a&=&N_f\frac{g_a^2}{16\pi^2} I_a(\Psi) \Lambda_{G}[\Psi] \nonumber \\
\Delta m_i^2&=&2\sum_aN_f\left(\frac{g_a^2}{16\pi^2}\right)^2C_a(i) I_a(\Psi)\Lambda_{S}^2[\Psi]
\end{eqnarray}
$\Lambda_G[\Psi]$ and $\Lambda_S^2[\Psi]$ are model dependent. In the limit $\frac{F}{M^2}<<1$ they are given by
\begin{eqnarray}
\Lambda_{G}[\Psi] &=& F_X\frac{\partial_X \mathcal M_\Psi}{\mathcal M_\Psi}\nonumber \\
\Lambda_{S}^2[\Psi] &=& |F_X|^2\left|\frac{\partial_X \mathcal M_\Psi}{\mathcal M_\Psi}\right|^2~,
\end{eqnarray}
where $X$ is the SUSY breaking spurion, $F_X$ is its $F$ component, and $M_\Psi$ is the fermion mass of the messenger $\Psi$. 

Three comments are in order.
The current version of \HSSOFTSUSY~does not support running hidden field masses. The masses are fixed, for the technical reason that they determine the different threshold scales.
Second, note that although there might be large splittings between the boson and fermion masses, 
the whole messenger multiplet is integrated out at the fermion mass scale.
The corrections due to this mass splitting were not implemented in this version of \HSSOFTSUSY.
The last comment is that users of \HSSOFTSUSY~who do not wish to discuss multiple messenger scales and doublet-triplet splittings are not required to specify all the details regarding the messenger sector, and are only required to provide the regular \SOFTSUSY~boundary conditions.

\section{Calculation algorithm}
The RG flow is determined by a complicated set of coupled differential equations.
Part of the boundary conditions (the measured standard model parameters) are given at the low scale boundary, and the others (user defined soft \SUSYB~terms) are given at the high energy boundary.
In addition to that, the system is constrained by the conditions of EW symmetry breaking at an intermediate scale (with a user defined $\tan\beta$).
In order to match between these constraints \SOFTSUSY~ implements an iterative procedure 
in which the set of differential equations is evolved back and forth between the different scales.
The details regarding the algorithm can be found in \cite{Allanach:2001kg}.
All the data and computation methods are held in a class named \code{MSSMSoftsusy}.

In this extension package a new class \code{HidSecSoftsusy} is defined.
It inherits all the members and methods of \code{MSSMSoftsusy}, and introduces the contributions of the hidden sector through the following intervention points:
\begin{itemize}
\item The iterative running between the different boundaries is divided into steps which are determined by the masses of the hidden fields. 

\item
At each step hidden fields are integrated out (or in),
and threshold effects are taken into account. The only threshold effects which are already 
implemented in \HSSOFTSUSY~are due to integration out of messengers (eq. (\ref{eq_threshold})).
Additional threshold effects can be supplied by the user via user defined functions.

\item
The contribution to the beta functions (eqs. (\ref{eq_beta1},\ref{eq_beta2},\ref{eq_beta3})) from  dynamical hidden fields (fields whose mass are smaller than the current energy scale) are taken into account automatically.
Additional contributions to beta functions can be supplied by the user via user defined functions.

\item
In the RGE engine of \SOFTSUSY~all the MSSM parameters are held in one array which is modified step by step along the RG flow according to the beta functions. If the hidden sector contains additional parameters which are scale dependent it is easy to include the new parameters in the RG evolution. This is done in the software by increasing the length of the array and holding the new parameters in the last cells.
The beta functions for these parameters should be supplied by the user via user defined functions.

\end{itemize}

\section{User defined hidden sector}
\HSSOFTSUSY~ allows two levels of intervention by the user -- defining the hidden sector fields and parameters in the input file, and by writing user defined C++ functions which modify the beta functions and threshold effects. The former does not require any programming, while the latter requires 
some basic understanding of the data structure in the software and several lines of additional code. 

\subsection{Input file format}
The input for \HSSOFTSUSY~ can be provided in a file written in what is known as the Les Houches Accord
format~\cite{Skands:2003cj,Allanach:2008qq}. In order to add a hidden field, a block in the following format should be added
\begin{verbatim}
Block HIDFIELD               # SU2 doublet messenger
    1   1.000000000e+10	     # Mass 
    2   1.000000000e+08      # Super Trace of the mass squared (only for messengers)
    3   1.000000000e+00	     # Number of flavors         
    4   0.500000000e+00      # Hypercharge
    5   2.000000000e+00	     # SU2 rep (SINGLET=1, FUND=2, ANTIFUND=3, ADJOINT=4)
    6   1.000000000e+00	     # SU3 rep (SINGLET=1, FUND=2, ANTIFUND=3, ADJOINT=4)
    7   2.000000000e+05      # LambdaG
    8   4.000000000e+10      # LambdaS squared    
    0   0.0                  # End of HidField Block  
\end{verbatim}
The data in the \code{HIDFIELD} block is given according to the following:
\begin{itemize}

\item \code 1 - The fermionic mass of the superfield (in GeV).

\item \code 2 - Str$[\mathcal M^2]=\sum m_b^2 - 2 m_f^2$ (in GeV$^2$). This quantity is relevant for the 2-loop contribution to the beta function of the scalar masses (eq. (\ref{eq_beta3})). If the hidden field is not a messenger, the value of this quantity can be set to zero.
\item \code 3 - The number of flavors, which is originally defined for QCD, will be used here to denote the degeneracy, or the number of copies of this field, for any type of field.
\item \code 4 - The hypercharge is given by its numerical value.
\item \code {5,6} - The representations under SU(2) and SU(3) which are supported in \HSSOFTSUSY~appear in table \ref{tab_enum} and are represented by the numerical value specified there.
\item \code {7,8} - The soft \SUSYB~scales defined in eq. (\ref{eq_threshold}) are given in GeV ($\Lambda_G$ in GeV, and $\Lambda_S^2$ in GeV$^2$). If the hidden field is not a messenger, the value of these quantities should be set to zero.
\item \code 0 - Each \code{HIDFIELD} block must end with a line with index \code{0}.
 
\end{itemize}

The hidden sector running parameters are defined in an optional block in the following format
\begin{verbatim}
Block HIDSECPARS             # Input parameters
    1   3.000000000e+00	     # Number of HidSecPars
    2   1.000000000e-01	     # 
    2   1.000000000e-02	     #
    2   1.000000000e-03	     # 
\end{verbatim}
The first integer input number indicates the number of running parameters in the hidden sector,
and the next items in the list give their value at the boundary\footnote{The line which contains the number of hidden sector parameters is marked by the index \code{1} and lines listing the boundary values of these parameters should be marked by \code{2}.}.

\subsection{User defined beta-functions}
The user defined beta-function accepts a const pointer for the \code{HidSecSoftsusy} class, and an array of numbers containing the value of the beta functions at the current energy scale. 
The list of the physical parameters in this array is given in table \ref{tab_par_array}.
The values of the beta function can be modified according to the new physics introduced in the model. The beta functions for the hidden sector parameters appear at the end of the array, at the same order as defined in the input file\footnote{In order to read the MSSM parameters from the \code{HidSecSoftsusy} class one can use the output member functions of \code{MSSMSoftsusy}, or use the member function \code{display()} which returns an array of parameters which are organized at the same order appearing in table \ref{tab_par_array}.}.
\begin{verbatim}
void (*userDefinedHidSecBeta)    (const HidSecSoftsusy *, DoubleVector &);
\end{verbatim}
\subsection{User defined threshold effect}
As we go down the RG flow, and the scale defined by the mass of a hidden field is reached, the theory is replaced by an effective theory where the hidden field is integrated out. Due to the integration out of the heavy modes, terms in the Lagrangian are generated or modified. In a similar manner, when we go up the RG flow,
and cross the mass threshold, the hidden field has to be integrated in. In \HSSOFTSUSY~ the threshold effects must be written explicitly in a user defined function. These functions (one for integrating out and one for integrating in) accept as input a pointer to the \code{HidSecSoftsusy} class, and the index of the field being integrated in/out\footnote{Note that in \code{HidSecSoftsusy} the hidden fields are held in an increasing mass order.}. Unlike the user-defined-beta-functions, in the threshold-effect-function the user can modify the MSSM parameters themselves, not only the beta functions, and therefore extra care is necessary when introducing changes.
\begin{verbatim}
void (*userDefinedIntegrateOutHidField)(HidSecSoftsusy *, int);
void (*userDefinedIntegrateInHidField) (HidSecSoftsusy *, int);
\end{verbatim}
\section{Running \HSSOFTSUSY}
\label{sec:run}
In order to use \HSSOFTSUSY~ the user must have a version of \SOFTSUSY~ installed.
The instructions for the installation can be found on the project's homepage. The \HSSOFTSUSY~ 
files should be copied into the project's directory\footnote{The files \code{Makefile.am} and \code{Makefile.in} should be overwritten.}, and the project should be recompiled:
\small
\begin{verbatim}
 > ./configure
 > make
\end{verbatim}
\normalsize
\HSSOFTSUSY~produces an executable called \code{hidsecsoftsusy.x} which accepts input in the SUSY Les Houches Accord 2 (SLHA2)~\cite{Allanach:2008qq}  input/output
option. The user must provide a file (\textit{e.g.} the example files included
in the \HSSOFTSUSY~distribution \code{HidSecInputSU5\_adjoint} and \code{HidSecInput\_EOGM}), 
that specifies the model dependent input parameters. The code may then be run with
\small
\begin{verbatim}
 > ./hidsecsoftsusy.x < input_file_name
\end{verbatim}
\normalsize

The output of the provided executable is based on the \SOFTSUSY~output in the SLHA2 format, followed by a list of the hidden fields with their mass and the hidden sector parameters with their values at the low energy boundary.

\section{\HSSOFTSUSY~usage examples}
In this section we describe three scenarios where \HSSOFTSUSY~can be used, and the required modification in the input file and software.
\subsection{Charged pseudo-moduli}
In the model described in \cite{Zur:2008zg} the hidden sector contains a TeV scale superfield in the adjoint representation of $SU(5)$ , which decomposes in the following way under $SU(3)\times SU(2)_L \times U(1)_Y$
\begin{equation}
 \label{decompose}
 {\bf 24}=({\bf 8},{\bf 1})_0\oplus({\bf 1},{\bf 3})_0\oplus
 ({\bf 3},{\bf2})_{-5/6}\oplus({\bf \bar 3},{\bf2})_{5/6}\oplus ({\bf 1},{\bf 1})_0 \ .
\end{equation}
In order to add these fields to the \HSSOFTSUSY~running, the input file should contain the following blocks:
\small
\begin{verbatim}
 ####################################################################
 Block HIDFIELD              # SU3 Octet superfield
    1   5.000000000e+04	     # Mass
    2   0.000000000e+00      # Mass2STrace
    3	  1.000000000e+00	     # N Flavors     
    4   0.000000000e+00	     # Hypercharge
    5   1.000000000e+00	     # SU2 rep (SINGLET=1, FUND=2, ANTIFUND=3, ADJOINT=4)
    6   4.000000000e+00	     # SU3 rep (SINGLET=1, FUND=2, ANTIFUND=3, ADJOINT=4)
    0   0.0                  # End of HidField Block 
####################################################################
Block HIDFIELD               # SU2 Triplet superfield
    1   5.000000000e+04	     # Mass
    2   0.000000000e+00      # Mass2STrace    
    3	  1.000000000e+00	     # N Flavors        
    4   0.000000000e+00      # Hypercharge
    5   4.000000000e+00	     # SU2 rep (SINGLET=1, FUND=2, ANTIFUND=3, ADJOINT=4)
    6   1.000000000e+00	     # SU3 rep (SINGLET=1, FUND=2, ANTIFUND=3, ADJOINT=4)
    0   0.0                  # End of HidField Block 
####################################################################   
 Block HIDFIELD              # SU3xSU2 bifundamental superfield 
    1   5.000000000e+04	     # Mass
    2   0.000000000e+00      # Mass2STrace   
    3	  1.000000000e+00	     # N Flavors         
    4  -0.833333333e+00      # Hypercharge
    5   2.000000000e+00	     # SU2 rep (SINGLET=1, FUND=2, ANTIFUND=3, ADJOINT=4)
    6   2.000000000e+00	     # SU3 rep (SINGLET=1, FUND=2, ANTIFUND=3, ADJOINT=4)
    0   0.0                  # End of HidField Block
####################################################################   
Block HIDFIELD               # SU3xSU2 bifundamental superfield 
    1   5.000000000e+04	     # Mass
    2   0.000000000e+00      # Mass2STrace   
    3	  1.000000000e+00	     # N Flavors         
    4   0.833333333e+00      # Hypercharge
    5   2.000000000e+00	     # SU2 rep (SINGLET=1, FUND=2, ANTIFUND=3, ADJOINT=4)
    6   3.000000000e+00	     # SU3 rep (SINGLET=1, FUND=2, ANTIFUND=3, ADJOINT=4)
    0   0.0                  # End of HidField Block 
####################################################################   
Block HIDFIELD               # singlet superfield (part of SU(5) adjoint)
    1   5.000000000e+04	     # Mass
    2   0.000000000e+00      # Mass2STrace   
    3	  1.000000000e+00	     # N Flavors         
    4   0.000000000e+00      # Hypercharge
    5   1.000000000e+00	     # SU2 rep (SINGLET=1, FUND=2, ANTIFUND=3, ADJOINT=4)
    6   1.000000000e+00	     # SU3 rep (SINGLET=1, FUND=2, ANTIFUND=3, ADJOINT=4)
    0   0.0                  # End of HidField Block 
\end{verbatim}
\normalsize

\subsection{EOGM}
In models with multiple messenger scales the \code{MODSEL} block in the input file should be set to \code{4}
\small
\begin{verbatim}
Block MODSEL		     # Select model
    1    4		     # sugra=1, gmsb=2, amsb=3, eogm=4
\end{verbatim}
\normalsize
and the \code{MINPAR} should not contain data regarding the messenger and SUSY breaking scales:
\small
\begin{verbatim}
Block MINPAR		     # Input parameters
    1   0.000000000e+00	     # m gravitino
    3   1.000000000e+01	     # tanb
    4   1.000000000e+00	     # sign(mu)
\end{verbatim}
\normalsize
The blocks defining the various messengers as hidden fields should contain the information regarding the scale of soft \SUSYB~terms generated when they are integrated out, and the supertrace of their mass squared: 
\small
\begin{verbatim}
Block HIDFIELD               # SU2 doublet messenger
    1   1.000000000e+10	     # Mass
    2   1.000000000e+08      # Mass2STrace   
    3	  1.000000000e+00	     # N Flavors         
    4  -0.500000000e+00      # Hypercharge
    5   2.000000000e+00	     # SU2 rep (SINGLET=1, FUND=2, ANTIFUND=3, ADJOINT=4)
    6   1.000000000e+00	     # SU3 rep (SINGLET=1, FUND=2, ANTIFUND=3, ADJOINT=4)
    7   2.000000000e+05      # LambdaG
    8   4.000000000e+10      # LambdaS2    
    0   0.0                  # End of HidField Block 
\end{verbatim}
\normalsize

\subsection{Running hidden sector parameters}
Consider a model where all the light hidden sector fields are charged under a $U(1)_D$ gauge group, which is weakly mixed with the $U(1)_{em}$ (like in theories of hidden dark sector e.g. \cite{ArkaniHamed:2008qn}) and assume the following beta function for  the coupling constant:
\begin{equation}
\beta_{g_D}=\sum_i(N_f)_i\frac{g_D^3}{16\pi^2}
\end{equation}
where the sum goes over the dynamical fields.
The coupling constant and its value at the UV boundary should be introduced into the input file by adding the following block:
\small
\begin{verbatim}
Block HIDSECPARS             # Input parameters
    1   1.000000000e+00	     # Number of HidSecPars
    2   1.000000000e-03	     # 
\end{verbatim}
\normalsize
In order to use the RGE engine of \SOFTSUSY~in determining the low energy value of the gauge coupling, the following function can be added to the software
\small
\begin{verbatim}
void MYuserDefinedHidSecBeta    (const HidSecSoftsusy * h, DoubleVector & d)
{
   
   //The current energy scale
   double ThisMu = h->displayMu();
   
   //The current value of the U(1)_D coupling in the beta function array
   double gD =  h->displayHidSecPar(1);
      
   double gDbetaValue = 0;
   const double oneO16Pisq = 1.0 / (16.0 * sqr(PI));
   
   //Adding the contributions of all the dynamical hidden sector fields
   for (int i = 1 ; i <= h->displayNHidFields() ; i++) 
   {
     if (ThisMu > h->displayHidFieldMass(i)) 
     {
	        gDbetaValue += oneO16Pisq * (gD*gD*gD)
	                         * h->displayHidFieldRepDimension(i,SU2)
	                         * h->displayHidFieldRepDimension(i,SU3) 
	                         * h->displayHidFieldNFlavors(i);
     }
   }
   
   //The index of the U(1)_D coupling in the beta function array
   int    gDIndex = numSoftParsMssm + 1;    
   
   //Modifying the beta function array accordingly
   d.set(gDIndex , gDbetaValue);
}
\end{verbatim}
\normalsize

\section*{Acknowledgments}
This software was written during research work with Y. Oz and L. Mazucatto.

\appendix

\section{Object Structure\label{sec:objects}}

The class in the \SOFTSUSY~ package in which the RG flow computation is implemented is named
\code{MSSMSoftsusy}. From the class \code{RGE} it inherits the engine which performs the 
iterative algorithm for solving the coupled differential equations, and it also contains all the
data regarding the MSSM parameters. In \HSSOFTSUSY~ we defined two new classes. The first, named \code{HidField}, holds the relevant information regarding a single field in the hidden sector. The second class, \code{HidSecSoftsusy}, inherits all the properties of \code{MSSMSoftsusy}, with an additional array of \code{HidField} objects, and a vector of numbers which can contain the hidden sector parameters.

\subsection{Class HidField}
\label{sec_HidField}
This class contains the physical parameters which define the field in the hidden sector. The list of parameters appears in table \ref{tab_HidField}. The gauge group and representation are defined using \code{enum} types \code{MSSMGaugeGroup} and \code{HidFieldRep} which appear in table \ref{tab_enum}.
In addition to the IO methods, it contains methods which compute the Casimir, the Dynkin index and the dimension of the representation under the 3 standard model gauge groups.

\begin{table}
\begin{center}
\begin{tabular}{lll}
  \textbf{data variable} & & \textbf{methods} \\ \hline
  \code{\small Double Mass} & Fermionic mass &
  \code{\small setMass}, \code{\small displayMass}
 \\
  $M$ [GeV] &  & 
  \\ \hline
    \code{\small Double HyperCharge} & Hypercharge and representation&
  \code{\small setSU2Rep}, \code{\small setSU3Rep}, \code{\small setHyperCharge},\\
  \code{\small HidFieldRep	SU2Rep, SU3Rep} &  under SU(2) and SU(3)  &  \code{\small displayHyperCharge},\\ 
  &  & \code{\small displaySU2Rep}, \code{\small displaySU3Rep}, \\
   &  & \code{\small displayCasimir}, \code{\small displayDynkinIndex}, \\
   & &  \code{\small displayRepDimension}   \\   \hline
  \code{\small int Nflavors} & Number of flavors &
  \code{\small setNflavors}, \code{\small displayNflavors}\\
  $N_f$ &  & 
  \\ \hline
  \code{\small Double LambdaG}, \code{\small LambdaS2} & Gaugino and scalar mass scales &
  \code{\small setLambdaG}, \code{\small setLambdaS2}
 \\
  $\Lambda_G$ [GeV], $\Lambda_S^2$ [GeV$^2$]& generated by integrating out this field. & 
  \code{\small displayLambdaG}, \code{\small displayLambdaS2}
  \\ \hline
  \code{\small Double StrM2} & Supertrace of the mass matrix. &
  \code{\small setStrM2}, \code{\small displayStrM2}
 \\
  Str$\{\mathcal{M}^2\}$ [GeV$^2$]&  & 

  \\ \hline
   \normalsize
\end{tabular}\end{center}
\caption{\code{HidField} class data and I/O methods.\label{tab_HidField}}
\end{table}

\begin{table}
\begin{center}
\begin{tabular}{llll|llll}
  \textbf{enum type} & & \textbf{Identifier} & \textbf{Group} & \textbf{enum type} & &\textbf{Identifier} & \textbf{Representation} \\ \hline
  
   \code{MSSMGaugeGroup} & 1&\code{U1} & $U(1)$ & \code{HidFieldRep} & 1&\code{SINGLET} & Singlet \\
                        & 2&\code{SU2} & $SU(2)$ &&2&\code{FUND} & Fundamental  \\
                        & 3&\code{SU3} & $SU(3)$&&3&\code{ANTIFUND} & Anti-fundamental  \\
                        &&&&&4&\code{ADJOINT} & Adjoint 
  \\ \hline
   \normalsize
\end{tabular}\end{center}
\caption{\code{MSSMGaugeGroup} and \code{HidFieldRep} enum types.\label{tab_enum}}
\end{table}

\subsection{Class HidSecSoftsusy}
\label{sec_HidSecSoftsusy}
\code{HidSecSoftsusy} is a derived class of \code{MSSMSoftsusy}, the class which implements the
MSSM spectrum calculation in \SOFTSUSY. In addition to the MSSM information it contains arrays of
hidden fields and hidden sector parameters, and pointers to user defined functions. The class contains the following member functions:

\subsubsection{\code{void addHidField (const HidField * newHidField)}}
This method accepts a pointer to \code{HidField} and copies its content into the \code{HidFields} array, maintaining an order by increasing mass.

\subsubsection{\code{int run  (double xi, double xf, double eps)}}
The \code{run} method of \code{MSSMSoftsusy} which implements the RG flow between two energy scales
is overloaded. In the \code{HidSecSoftsusy} version of the method the running is divided into steps which are determined by the masses of the hidden fields. At each step the method \code{IntegrateOutHidField} (or \code{IntegrateInHidField}) is called.

\subsubsection{\code{void IntegrateOutHidField (int i), IntegrateInHidField (int i)}}
This method calls the user defined function for integrating out/in a hidden sector field.
If the hidden sector field is a messenger (it has non-zero $\Lambda_G$ or $\Lambda_S$), it also calls \code{AddOrRemoveMessenger} which implements eq. (\ref{eq_threshold}). Users who wish to provide the user defined function should note that the serial number of the hidden sector, \code{i}, is determined by its mass.

\subsubsection{\code{DoubleVector beta() const}}
The \code{beta} method of \code{MSSMSoftsusy} which computes the derivative of the MSSM couplings
is modified to include the hidden sector contributions discussed in sec. \ref{sec_HidSeccontribution}. It also calls the user defined beta function. 
The output is held in a \code{DoubleVector} array. Users who wish to provide the user defined beta function can use table \ref{tab_par_array} which lists the location of the various physical parameters in this array.

\subsubsection{\code{bool AnomalyCancellation()}}
This method checks whether the hidden fields satisfy the anomaly cancellation conditions.
\begin{table}
\begin{center}
\begin{tabular}{ll|ll}
\# & \textbf{Physical parameter} &\# &\textbf{Physical parameter} \\ \hline
  $1-9$ & $u$ quark Yukawa couplings &    $37-45$ & $u$-Higgs quark trililnear couplings \\  
  $10-18$& $d$ quark Yukawa couplings&    $46-54$ & $d$-Higgs quark trililnear couplings \\  
  $19-27$& lepton Yukawa couplings   &    $55-63$ & lepton-Higgs trililnear couplings \\  
 $28$ & $U(1)~$ gauge coupling       &    $64-72$ & quark doublet scalar mass squared \\
 $29$ & $SU(2)$ gauge coupling       &    $73-81$ & $\bar u$ quark scalar mass squared \\
 $30$ & $SU(3)$ gauge coupling       &    $82-90$ & $\bar d$ quark scalar mass squared \\
 $31$ & $\mu$                        &    $91-99$ & lepton doublet scalar mass squared \\
 $32$ & $\tan\beta$                  &    $100-108$ &$\bar e$ doublet scalar mass squared \\
 $33$ & The Higgs VEV                &    $109$     & $B\mu$                             \\
 $34$    &$U(1)~$ gaugino mass       &    $110$     & $m_{Hd}^2$                            \\
 $35$    &$SU(2)~$ gaugino mass      &    $111$     & $m_{Hu}^2$                            \\
 $36$    &$SU(3)~$ gaugino mass      &    $112-\ldots$ & hidden sector parameters     \\
  \hline
   \normalsize
\end{tabular}\end{center}
\caption{The order of physical parameters in the beta function array.\label{tab_par_array}}
\end{table}

\end{document}